# Brexit or Bremain ?
# Evidence from bubble analysis

**Marco Bianchetti**, Intesa Sanpaolo, University of Bologna[1]
**Davide Galli**, Università degli Studi di Milano, Physics Dept.
**Camilla Ricci**
**Angelo Salvatori**, Università degli Studi di Milano, Physics Dept.
**Marco Scaringi**, Università degli Studi di Milano, Physics Dept.

**Abstract**

We applied the Johansen-Ledoit-Sornette (JLS) model to detect possible bubbles and crashes related to the Brexit/Bremain referendum scheduled for 23rd June 2016. Our implementation includes an enhanced model calibration using Genetic Algorithms. We selected a few historical financial series sensitive to the Brexit/Bremain scenario, representative of multiple asset classes.
We found that equity and currency asset classes show no bubble signals, while rates, credit and real estate show super-exponential behaviour and instabilities typical of bubble regime. Our study suggests that, under the JLS model, equity and currency markets do not expect crashes or sharp rises following the referendum results. Instead, rates and credit markets consider the referendum a risky event, expecting either a Bremain scenario or a Brexit scenario edulcorated by central banks intervention. In the case of real estate, a crash is expected, but its relationship with the referendum results is unclear.

---

[1] Corresponding author, marco.bianchetti(at)unibo.it.

## 1. Brexit or Bremain ?

On Dec. 17, 2015 the UK Parliament approved the European Union Referendum Act 2015 to hold a referendum on whether the United Kingdom should remain a member of the European Union (EU). The referendum will be held[2] on Jun. 23, 2016, with the following Q&A:
- Q: "Should the United Kingdom remain a member of the European Union or leave the European Union?
- A1: "Remain a member of the European Union"
- A2: "Leave the European Union"

The two scenarios above were called "Bremain" and "Brexit", respectively. In case of Brexit decision, there is no immediate withdrawal. Instead, a negotiation period begins to establish the future relationship between UK and EU. The negotiation length is two years, extendible upon agreement between the two parties. For example, the agreements between EU and Switzerland took 10 years of negotiations.

Referendum campaigning has been suspended on 16th June 2016 following the shooting of Labour MP Jo Cox. This event has had a strong impact on the public opinion, rapidly changing the opinion polls and possibly the attitude of the country.

Forecasting the results of the 23[rd] June 2016 referendum, given the apparent parity between Bremain and Brexit supporters and the high percentage of undecided voters observed until the week before, is clearly a very challenging task, with a high error probability. Nevertheless, there exist at least three sources of data supporting forecast analysis: opinion polls [8] [10], bookmakers betting odds [9], and market data [10]. In this paper we recur to a different forecasting approach, described in the next section.

## 2. Methodology

We applied a forecasting methodology based on the Johansen-Ledoit-Sornette (JLS) model, developed since the 90s at ETHZ by D. Sornette and co-authors (see e.g. [1]-[4] and refs. therein). The JLS model has been extensively applied to bubbles, crashes and crisis analysis in many fields. For applications in finance see e.g. the Financial Crisis Observatory [5].

The JLS model assumes that, during a bubble regime, the asset mean value follows the so-called Log-Periodic Power Law (LPPL) function,

$$LPPL(t) = A + B(t_c - t)^m + C(t_c - t)^m cos(\omega \, log(t_c - t) + \phi),$$

---
[2] We stress that this paper was delivered before the UK referendum scheduled for 23rd June 2016.

$$LPPL(t) = ln[\mathbb{E}_t[p(T)]] = ln[p(t)], \qquad (1)$$

where $p(t)$ is the asset price and $\mathbb{E}_t[p(T)]$ denotes the conditional expectation of the future value $p(T)$ at present time $t < T$, given all information available up to time $t$. In eq. (1) above, $A$ is the value $ln[p(t_c)]$ at the critical time, $B < 0$ is the increase in $ln[p(t)]$ over the time unit before the crash if $C$ were to be close to zero, $0 < m < 1$ should be positive to ensure a finite price at the critical time $t_c$ and lower than one to quantify the super-exponential acceleration of price $p(t)$, $C \neq 0$ is the proportional magnitude of the oscillations around the exponential growth $\omega$ is the frequency of the oscillations during the bubble, and finally $0 < \phi < 2\pi$ is a phase factor. Note that the seven JLS parameters $\{A, B, C, m, \omega, \phi, t_c\}$ are all free parameters that must be calibrated to fit the asset's historical series, without imposing a known critical time $t_c$. Extensive backtesting of the JLS model on past bubbles allowed to identify more stringent parameters constraints, namely $0.1 < m < 0.9$, $6 < \omega < 13$, and $|C| < 1$ [3].

Overall, the JLS model describes the dynamics of a system with a growing instability, generated by behaviors of investors and traders that create positive feedback in the valuation of assets leading to unsustainable growth and culminating with a finite-time singularity at some future critical time $t_c$, which is interpreted as the forecast of a possible crash. A voluminous literature has applied this model (and slightly different versions) to various financial data, detecting many historical cases to which the log-periodic apparatus could be applied. We refer the reader to [1]-[4] and to references therein for more details.

Our implementation of the JLS model is based on the original version [1]-[4], enhanced with robust global optimization methods, i.e. Genetic Algorithms, for model calibration [6]. The JLS model calibration requires the optimal fit of the historical series with the LPPL function. The fit is optimal if the set $\wp = \{A, B, C, m, \omega, \phi, t_c\}$ of LPPL parameters minimizes the root mean square error between the historical series and the LPPL fit function, $RMS(\wp)^2 = \sum_{i=1}^{N}[p(t_i) - LPPL(t_i, \wp)]^2$, where $\{t_1 \cdots t_N\}$ and $\{p_1 \cdots p_N\}$ are the historical dates and prices, respectively.

The calibration problem above is computationally hard, since the oscillating term in the LPPL function produces many local minima in the RMS error function, where the minimization algorithm get trapped. This is the reason why different calibration strategies have been proposed in the literature [3]. Our global optimization approach attacks the problem without any assumption on the shape of the LPPL hyper-surface[3] and ensures to find the global minimum corresponding to the optimal fit. Clearly, such approach is much more computationally demanding, and required appropriate parallel computing facilities [7].

We applied the JLS model to a selection of historical financial series sensitive to the current Brexit/Bremain scenario. For each series, we run multiple model calibrations with different calibration windows, and detected possible bubble signals, corresponding to possible critical times $t_c$. Such bubble signals were accepted or rejected according to the constraint discussed above. This procedure ensures the stability of the observed results.

---

[3] In particular, we do not use taboo search or multiple local optimizations as described in [3].

## 3. Results

The results are reported in the following Figure 1- Figure 8. The description of the market data and the comments are included in their corresponding captions. Each chart shows, on the left hands scale, the historical series (blue line), and one representative fit with LPPL function in eq. (1) (red line) among many calibrations run with different calibration windows. The histograms reported on the right hand scale count the bubble signals (if any). In case of no bubble signals, no histograms appear.

The interpretation of the occurrence or not of the JLS bubble signal deserves some attention. The theory behind the JLS model states that if investors in some asset expect a future event (e.g. the UK Referendum) leading to a possible negative scenario for that asset (e.g. Brexit), this may trigger an asset dynamics leading to a bubble regime, possibly followed by a crash. Thus, reversing the argument, if one detects bubble signals for an asset and knows how a future event will affect the asset price, then one can state that the investors expect a negative scenario for that asset.

Translating into the Brexit context, if one detects bubble signals for an asset with a critical time $t_c$ around June 23th, and knows that Brexit/Bremain are negative/positive scenarios for that asset, respectively, one can conclude that investors are expecting Brexit. The specular argument also holds: if one knows that Bremain/Brexit are negative/positive scenarios for that asset, respectively, one can conclude that investors are expecting Bremain.

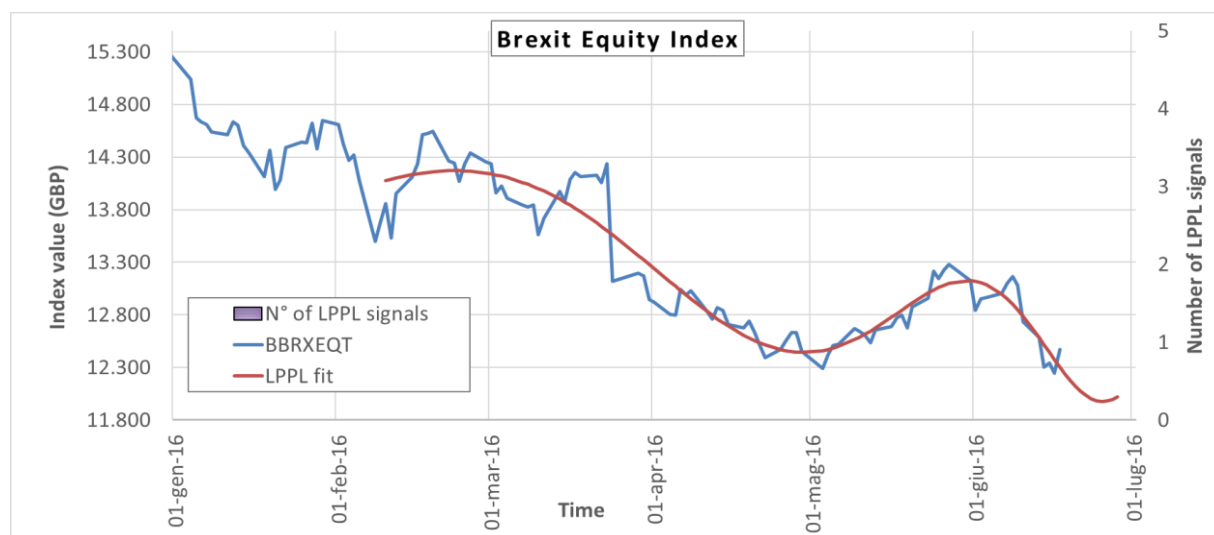

**Figure 1**

- **Source**: Brexit Equity Index (Bloomberg BBRXEQT Index), basket of 10 UK stocks designed to reflect British exposure to the EU across different sectors. Data up to Friday 17th June 2016.
- **Comments**: the historical series shows a decreasing trend, but no super-exponential behaviour and instabilities typical of bubble regime. In fact, the JLS model (LPPL fit) does not propose valid bubble and crash signals.
- **Interpretation**: market participants are currently suspicious about UK stock market, but do not actually fear either a crash following Brexit or a sharp rise following Bremain.

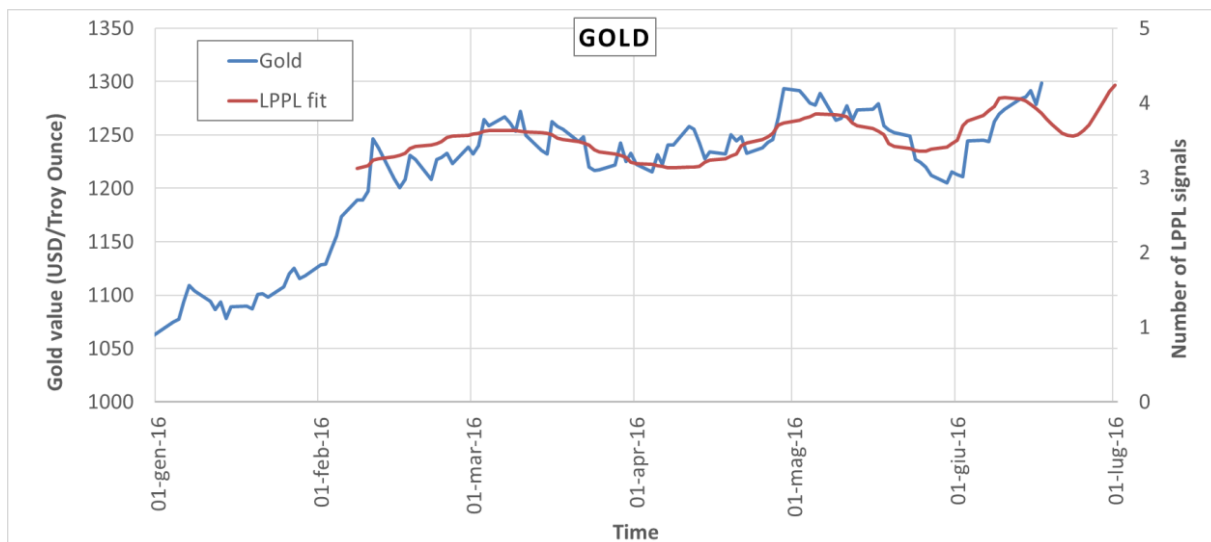

**Figure 2**

- **Source**: gold prices (Bloomberg XAU BGN Crncy). Data up to Friday 17th June 2016.
- **Comments**: the historical series shows an increasing trend, but no super-exponential behaviour and instabilities typical of bubble regime. In fact, the JLS model (LPPL fit) does not propose valid bubble and crash signals.
- **Interpretation**: market participants are currently refuging into gold, but do actually fear neither a sharp rise following Brexit nor a crash following Bremain. This result is consistent with the BBRXEQT and GBPUSD FX rate observations.

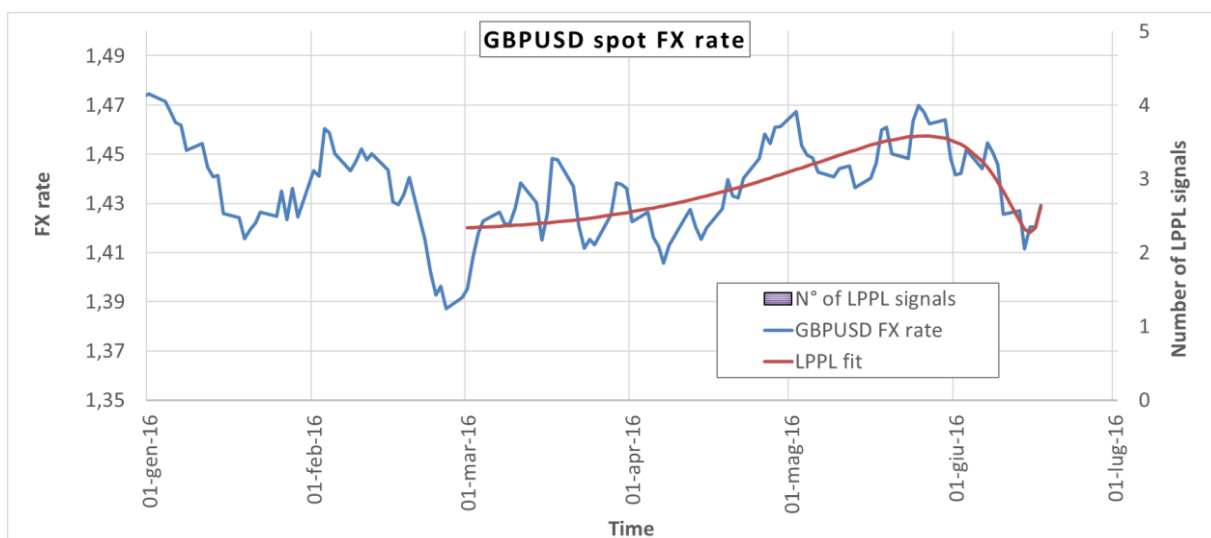

**Figure 3**

- **Source**: GBP/USD FX rate (Bloomberg GBPUSD BGN Crncy). Data up to Friday 17th June 2016.
- **Comments**: the historical series shows an erratic trend, no super-exponential behaviour and instabilities typical of bubble regime. In fact, the JLS model (LPPL fit) does not propose valid bubble and crash signals.
- **Interpretation**: market participants but do not actually fear either a crash following Brexit or a sharp rise following Bremain. This result is consistent with the BBRXEQT and GBPUSD FX rate observations.

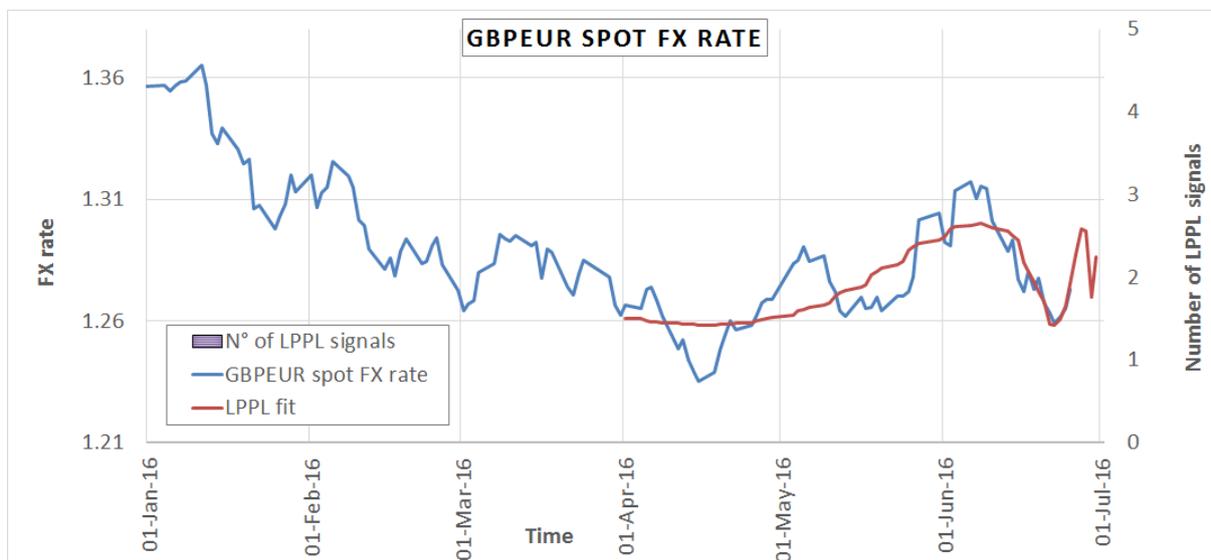

**Figure 4**

- **Source**: GBP/EUR FX rate (Bloomberg GBPEUR BGN Crncy). Data up to Friday 17th June 2016.
- **Comments**: as for GBP/USD
- **Interpretation**: as for GBP/USD.

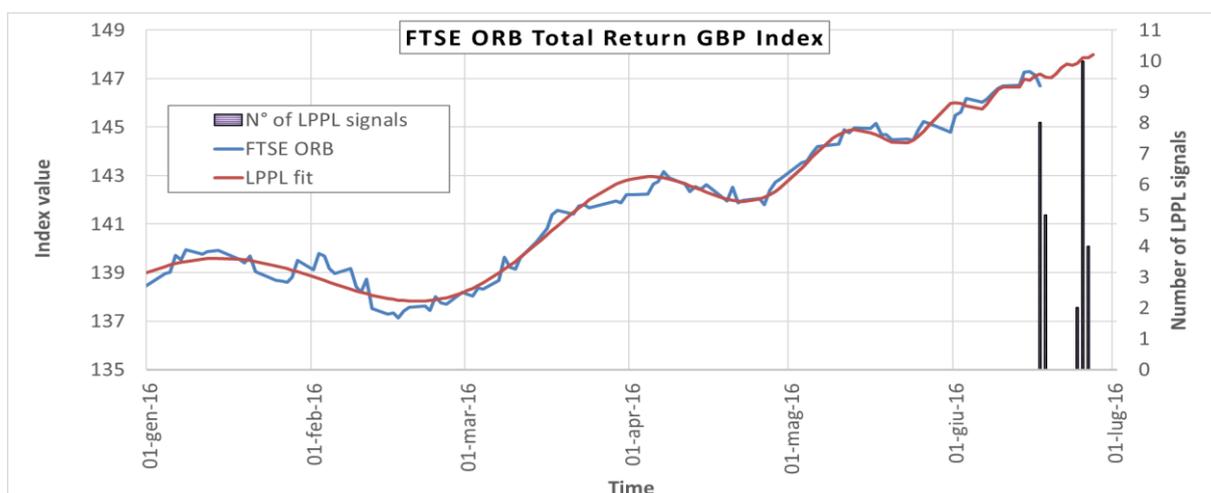

**Figure 5**

- **Source**: FTSE ORB Total Return GBP Index (Bloomberg TFTSEORB Index), includes GBP fixed coupon Corporate bonds trading on LSE across different industry sectors and maturity bands. Data up to Friday 17th June 2016.
- **Comments**: the historical series shows an upward trend (due to the overall lowering discount rates, driven by lowering GBPLibor w.r.t. increasing GBP credit spreads) and super-exponential growth and instabilities typical of bubble regime. In fact, the JLS model (LPPL fit) propose several valid crash signals around 23th June.
- **Interpretation**: market participants consider the referendum a risky event for corporate bonds, expecting either a Bremain scenario or the BoE intervention in case of Brexit.

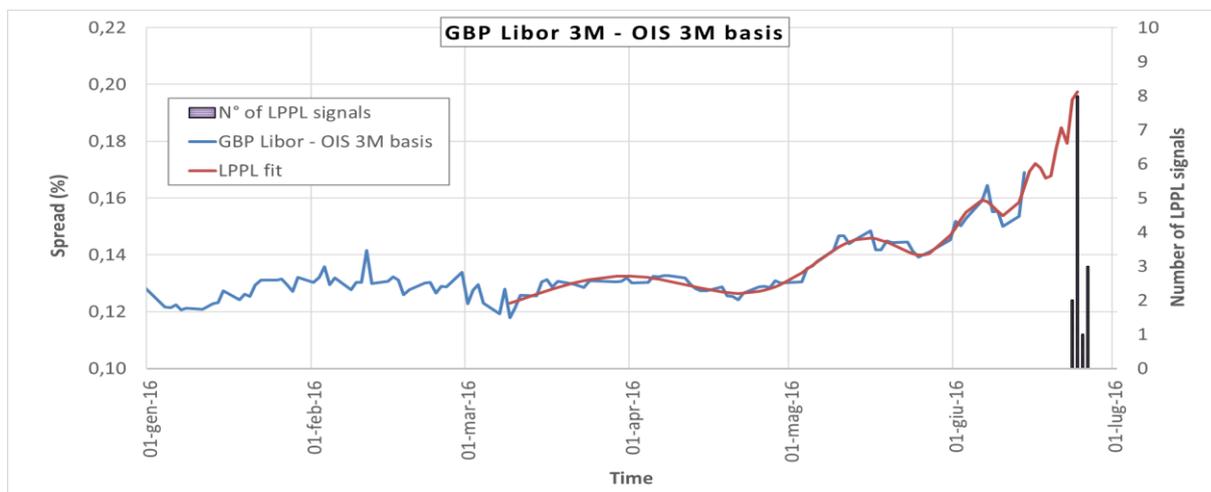

**Figure 6**

- **Source**: GBPLibor3M vs GBP OIS 3M (Bloomberg BP003M Index – BPSWSC Crncy). Measures the London interbank credit and liquidity risk on 3M time horizon relative to overnight horizon. Data up to Friday 17th June 2016.
- **Comments**: the historical series shows super-exponential behavior and instabilities typical of bubble regime. In fact, the JLS model (LPPL fit) does propose valid bubble and crash signals around 24th June.
- **Interpretation**: market participants expect that the basis spread will crash back to lower values, corresponding to lower credit and liquidity risk in the London interbank market. This result is consistent with the FTSE ORB observations.

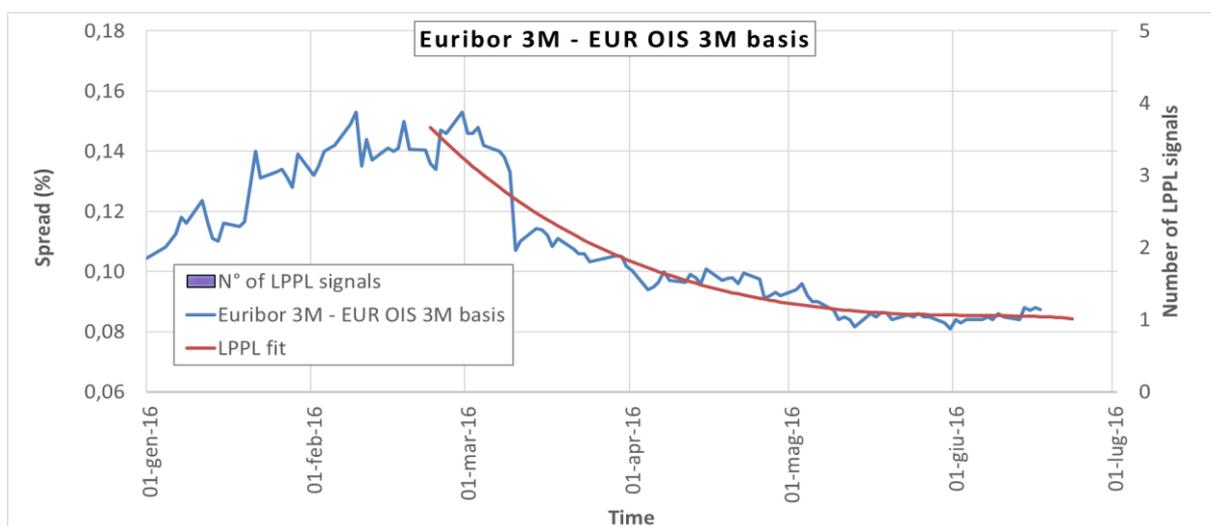

**Figure 7**

- **Source**: Euribor3M vs EUR OIS 3M (Bloomberg EUR003M Index – EUSWEC Crncy). Measures the EUR interbank credit and liquidity risk on 3M time horizon relative to overnight horizon. Data up to Thursday 16th June 2016.
- **Comments**: the historical series shows a decreasing trend but no super-exponential behaviour and instabilities typical of bubble regime. In fact, the JLS model (LPPL fit) does not propose valid bubble and crash signals.
- **Interpretation**: market participants but do not actually fear either a crash following Brexit, also because the expected ECB intervention, or a sharp rise following Bremain.

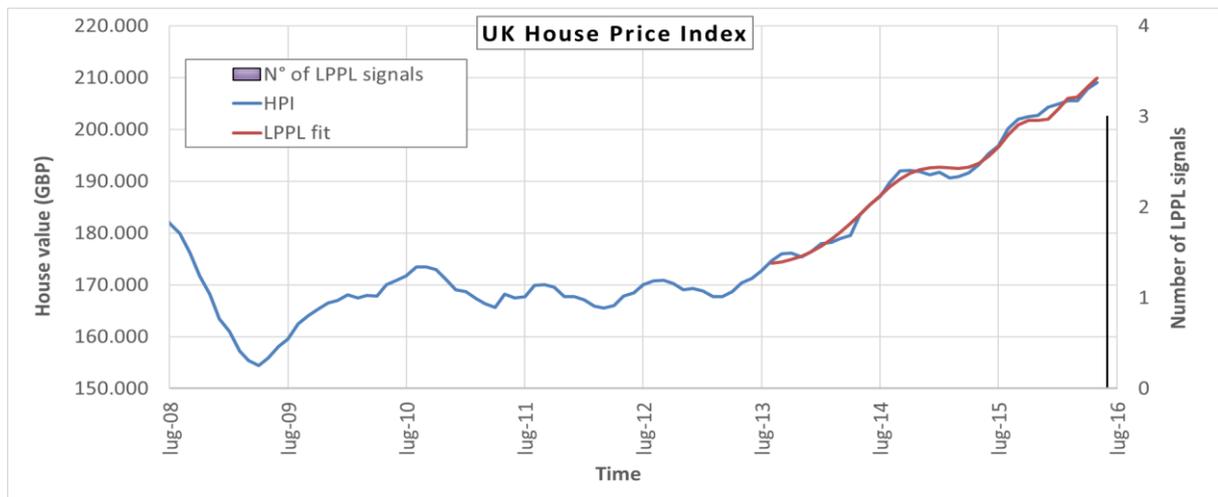

**Figure 8**

- **Source**: UK house price index 9. Data up to April 2016 (this data is updated with delay).
- **Comments**: the historical series shows an increasing trend with super-exponential behaviour and instabilities typical of bubble regime. In fact, the JLS model (LPPL fit) does propose valid bubble and crash signals around June.
- **Interpretation**: the trend remembers those observed during the 2008 subprime crisis. Market participants expect a crash, but its relationship with the referendum is questionable, since the growth regime started before the current Brexit/Bremain context, and more recent UK HPI data would be needed.

| # | Asset class | Historical series | JLS bubble signals |
|---|---|---|---|
| 1 | Equity | Bloomberg Brexit Equity Index | NO |
| 2 | | Gold | NO |
| 3 | Currency | GBPUSD Spot FX Rate | NO |
| 4 | | GBPEUR Spot FX Rate | NO |
| 5 | | FTSE ORB Total Return GBP Index | YES |
| 6 | Rates and credit | GBP Libor - GBP OIS 3M basis | YES |
| 7 | | Euribor - EUR OIS 3M basis | NO |
| 8 | Real estate | UK House Price Index | YES |

**Table 1**: summary of JLS bubble signals (col. 4) from Figure **1**- Figure **8**.

## 4. Conclusions

We applied a forecasting methodology based on the Johansen-Ledoit-Sornette (JLS) model, developed since the 90s by D. Sornette at ETHZ and co-authors [1][1], and extensively applied to detect bubbles, crashes and crisis in many fields [5]. Our implementation includes an enhanced model calibration using robust global optimization methods, i.e. Genetic Algorithms [6].

We applied the JLS model to a selection of historical financial series sensitive to the current Brexit/Bremain scenario, representative of equity (BBRXEQT), currency (Gold, GBPUSD and GBPEUR fx), rates and credit (FTSE ORB, GBP and EUR Libor – OIS basis), and real estate (UK HPI) asset classes.

We found the following evidence (see Table 1):

- equity and currency asset classes show no bubble signals,
- rates, credit and real estate show super-exponential behaviour and instabilities typical of bubble regime, with the exception of Euribor-EUR OIS basis.

Out study suggests that, under the JLS model, the following interpretations can be drawn:
- equity and currency: market participants coherently do not expect crashes or sharp rises following the referendum results.
- Rates and credit: market participants coherently consider the referendum a risky event for the London market, expecting either a Bremain scenario or a Brexit scenario edulcorated by central banks intervention.
- In the case of real estate, market participants expect a crash, but its relationship with the referendum results is unclear.

## 6. Disclaimer and acknowledgments

**Disclaimer**

The views and the opinions expressed in this document are those of the authors and do not represent the opinions of their employers. They are not responsible for any use that may be made of these contents. The opinions, forecasts or estimates included in this document strictly refer to the document date, and there is no guarantee that future results or events will be consistent with the present observations and considerations. This document is written for informative purposes only, it is not intended to influence any investment decisions or promote any product or service.

**Acknowledgments**
The authors gratefully acknowledge Luca Lopez for fruitful discussion and analysis at the early stage of this work.